\documentclass[fleqn,twoside,twocolumn,nofootinbib,showkeys]{revtex4} 
\usepackage[sec,nocpr]{ujp}
\usepackage[utf8]{inputenc}
\usepackage{graphicx}
\usepackage[colorlinks,hypertexnames=false]{hyperref}

\razd{\seci}

\begin{document}

\def\bb    #1{\hbox{\boldmath${#1}$}}

\def\qb{{\bar q}}
\def\qcdu{{{}_{ \rm { QCD}}}}   
\def\qedu{{{}_{\rm { QED}}}}   
\def\qcdd{{{}^{ \rm QCD}}}   
\def\qedd{{{}^{\rm QED}}}   
\def\qcd{{{\rm QCD}}}   
\def\qed{{{\rm QED}}}   
\def\2d{{{}_{\rm 2D}}}         
\def\4d{{{}_{\rm 4D}}}         
\def\sg#1{ {\rm \,sign}(#1)\, }

\title[Color-Singlet and Color-Octet $q\bar q$ Quark Matters]
{COLOR-SINGLET AND COLOR-OCTET QUARK MATTERS$^{1,2}$}%
\author{Cheuk-Yin Wong}
\affiliation{Physics Division, Oak Ridge National Laboratory, Oak Ridge, TN 37831 U.S.A.}
\email{wongc@ornl.gov}

\autorcol{Cheuk-Yin Wong} 

\cit{Cheuk-Yin Wong}

\setcounter{page}{1}

\begin{abstract}
Quarks and antiquarks carry color and electric charges and belong to
the color-triplet  $3$   group and the color-antitriplet ${\bar
  3}$ group respectively.  The product groups of ${3}$ and $
{\bar 3}$ consist of the color-singlet ${1}$ and the color-octet
${8}$ subgroups.  Therefore, quarks and antiquarks combine to form
color-singlet $[q \bar q]^1$ quark matter and color-octet $[q \bar
  q]^8$ quark matter.  The color-octet quark matter corresponds to the
$q\bar q$ quark matter as envisaged in the realm of present knowledge
but the color-singlet quark matter is as yet unexplored and now
submitted for exploration.  The color-singlet quark matter with two
flavors can be separated into charged and neutral color-singlet quark
matters.  In the neutral color-singlet quark matter, the quark and the
antiquark interacting only in the QED interaction may form stable and
confined colorless QED mesons non-perturbatively at about 17 MeV and
38 MeV (PRC81,064903(2010) and JHEP(2020(8),165).  It is proposed that
the possible existence of the QED mesons may be a signature of the
neutral color-singlet quark matter at $T=0$.  The observations of the
anomalous soft photons at CERN, and the anomalous bosons with mass
about 17 MeV at ATOMKI, DUBNA, and HUS, and mass about 38 MeV at DUBNA
hold promising experimental evidence for the existence of such QED
mesons, pending further confirmations.

\end{abstract}


\maketitle

\section{Introduction}
\setcounter{footnote}{1} \footnotetext {The research was supported in
  part by UT-Battelle, LLC, under contract DE-AC05-00OR22725 with the
  US Department of Energy (DOE). The US government retains the
  publisher, by accepting the article for publication, acknowledges
  that the US government retains a nonexclusive, paid-up, irrevocable,
  worldwide license to publish or reproduce the published form of this
  manuscript, or allow others to do so, for the US government
  purposes. DOE will provide public access to the results of federally
  sponsored research in accordance with the DOE Public Access Plan
  (http://energy.gov/downloads/doe-public-access-plan) }
\setcounter{footnote}{2} \footnotetext {Based on a talk presented at
  the International Conference on New Trends in High Energy Physics,
  Batumi, Georgia, September 15-19, 2025.}

It is fitting that we dedicate the Proceedings of this Conference to
the memory of Prof.\ V\'aszl\'o Jenkovszky, not only because he was
one of the founders of this series of conferences, but also because of
his many important contributions to high energy physics and his
promotion of friendship among physicists of all nations.  He firmly
believed that contacts and collaborations among physicists of
different nations would lead to better understanding among peoples of
all countries, and would promote the development of peace as well as
science and technology.  He will be well remembered by all those who
knew him.

In the realm of present knowledge, the material and the interaction in
the conventional "quark matter" has been generally considered to
consist of quarks interacting predominantly in the non-Abelian SU(3)
QCD interaction, with the Abelian U(1) QED interaction as a small
perturbation.  As a function of temperature $T$, the manifestation of
the quark matter in different phases has been well studied
\cite{Won93,Yag05,Vog07}.  The extension of the consideration of quark
matter to other unexplored color degrees of freedom in conjunction
with the nonperturbative inclusion of the U(1) QED interaction will
bring us to uncharted new frontiers.

Quarks and antiquarks carry color and electric charges and belong to
the color-triplet $\bb{3}$ group and the color-antitriplet $\bb{\bar
  3}$ group respectively.  From the group theoretical considerations,
we have $\bb{3} \, \otimes \, \bb{\bar 3} = \bb{1} \, \oplus \, \bb
{8}$.  That is, the direct product of the $\bb{3}$ group and the $\bb
{\bar 3}$ group consists of the color-singlet subgroup $\bb{1}$ and
the color-octet subgroup $\bb{8}$.  Therefore, quarks and antiquarks
combine to form the color-singlet $[q \bar q]^1$ quark matter and the
color-octet $[q \bar q]^8$ quark matter, where the superscripts denote
color-multiplet indexes.

By the principle of the colorless nature of observable entities,
observable $q\bar q$ quark matter complexes must be colorless
color-singlet entities. Consequently, to form color-singlet states of
these complexes in their lowest-energy regions at $T$=0, the
color-octet quark matter must interact first with the color-octet
SU(3) QCD interactions non-perturbatively to form QCD-confined
colorless $[[q\bar q]^8g^8]^1$ complexes.  Additional QED interaction
in the complexes can then be included either perturbatively or
nonperturbatively, and such an additive inclusion does not change the
colorless nature nor the confinement property of these colorless
complexes.  We therefore recognize the $q\bar q$ color-octet quark
matter as corresponding to the $q \bar q$ quark matter conventionally
envisaged in the realm of our present knowledge.

On the other hand, the color-singlet quark matter is as yet
unexplored.  We wish to study theoretically colorless physical
complexes composed of such interacting quark matter at $T$=0 as the
doorway states for future exploration of the different phases of the
neutral color-singlet quark matter at higher temperatures.  To form
color-singlet complexes of such quark matter in the lowest-energy
region at $T$=0, the color-singlet quark matter must interact only
with the color-singlet U(1) QED interaction.  Inclusion of the QCD
interaction in the color-singlet quark matter will change the
colorless nature of these color-singlet complexes, or would place them
to much higher energy excited states such as the $[[q\bar
    q]^1[gg]^1]^1$ states and the glueball states.  The color-singlet
quark matter interacting only in the Abelian U(1) QED interaction
brings us to a new sector of the $q\bar q$ quark matter in uncharted
territories.

We focus our attention mainly on $q\bar q$ quark matter with two light
flavors.  We assume isospin symmetry so that $I_z$ and $I$ are good
quantum numbers.  In such a sector, the color-singlet $q \bar q$ quark
matter can be separated into charged color-singlet quark matter with
$I$=$1$ and $I_z$=$\pm 1$ and neutral color-singlet quark matter with
$I_z$=$0$ and $I$=0,1 as depicted in Fig.\ 1.  We shall also examine
the $qqq$ quark matter with baryons and the QED neutron in Section
{\ref{qqq}.

\vspace {-0.2cm}
\begin{figure} [h]
  \centering
\hspace*{-0.0cm}%
\includegraphics[width=8.0cm,height=8.0cm]{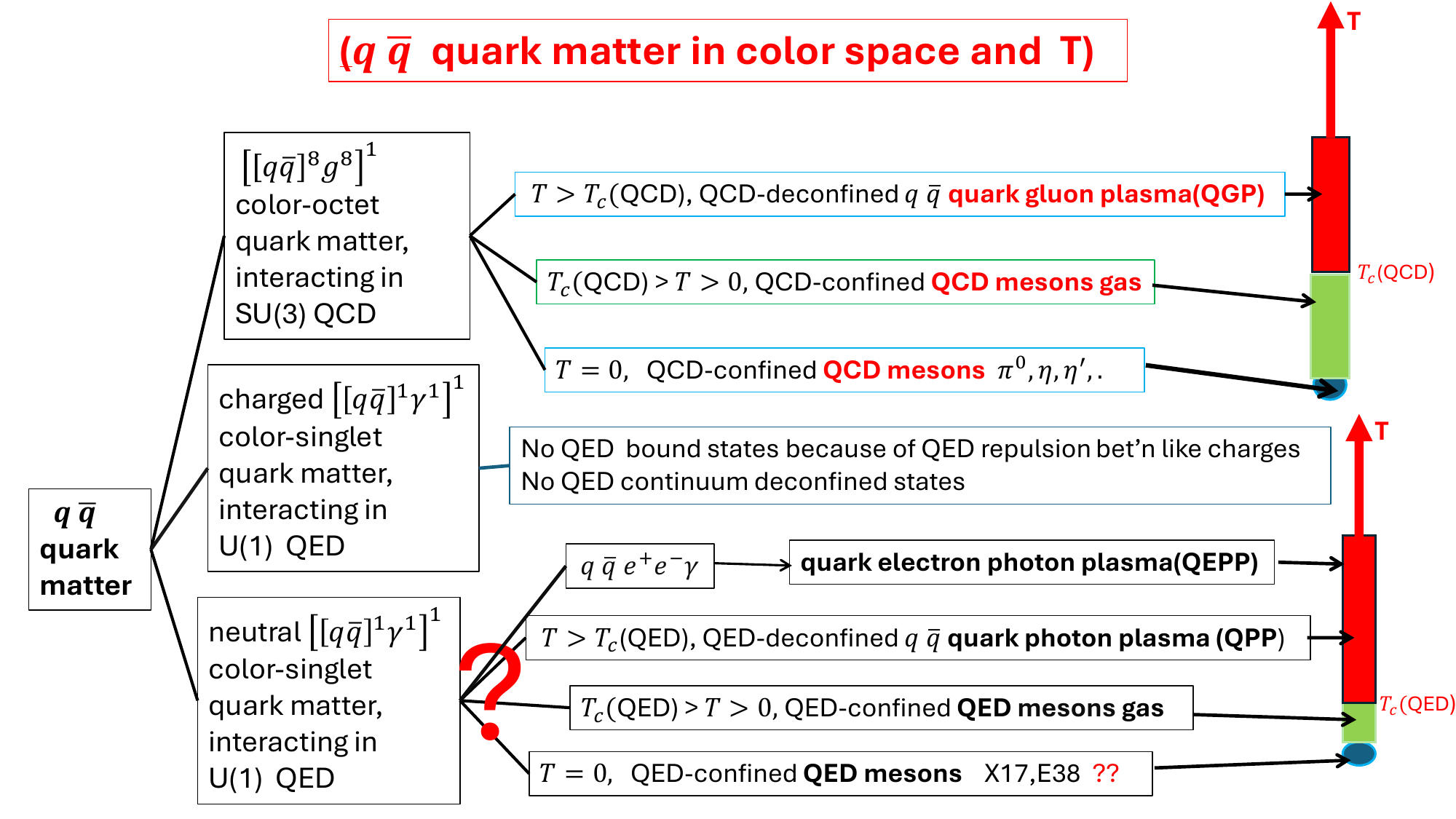}
\vspace*{-0.3cm}
\caption{The classification of the $q\bar q$ quark matter with two light flavors  in color space,  temperature $T$, 
and  possible  phases. The big question mark points to  new  and unexplored physics frontiers worthy of further exploration.}
\label{fig1}
\end{figure}

Charged color-singlet quark matter with $I$=1 and $I_z$=$\pm1$
involves $|u \bar d\rangle$ and $|d \bar u\rangle$ states with the
quark and the antiquark possessing electric charges of the same sign.
It will not possess stable $q\bar q$ bound states in the U(1) QED
interaction because of the repulsion between the quark and the
antiquark with electric charges of the same sign.  Furthermore, as
isolated quarks do not exist, the charged color-singlet quark matter
does not possess QED continuum unbound $q\bar q$ states.
Consequently, below the pion mass threshold, charged color-singlet
quark matter does not leave remarkable imprints of its existence at
$T$=0.  Above the pion mass threshold, the strongly attractive QCD
interaction in the quark matter dominates the confinement dynamics to
lead to charged QCD meson states in which the weaker repulsive QED
interaction gives only a perturbation to the QCD meson energy levels.

On the other hand, neutral color-singlet quark matter with $I_z$=$0,
I$=$0,1$ will involve a linear combination of $|u \bar u\rangle$ and
$|d \bar d\rangle$ states with the quark and the antiquark possessing
electric charges of opposite signs and may form stable and confined
states at $T$=0 because of the attractive QED interaction between the
quark and the antiquark and the Schwinger confinement mechanism.  The
QED-confined $q\bar q$ composite particles in neutral color-singlet
quark matter at $T=0$ can be called the QED mesons, in analogy with
the QCD mesons arising from the attractive color-octet SU(3)
interaction between a quark and an antiquark in QCD at $T=0$.  If QED
mesons exist at $T$=0, then, by further analogy with QCD, there will
likely be a critical temperature $T_c$(QED) below which an assembly of
neutral color-singlet $q\bar q$ quark matter would exist as a QED
meson gas.  Above $T_c$(QED), the color-singlet quark matter would
likely be deconfined quarks, antiquarks, and photons in a quark photon
plasma.  At a high temperature, when the photons split into
electron-positron pairs, the neutral color-singlet quark matter may
exist as a quark electron photon plasma.  The classification of quark
matter in color space, temperature $T$, and possible phases is
depicted in Fig. 1.

Previously, QED mesons consisting of $q$ and $\bar q$ interacting
non-perturbatively only in QED were predicted to have masses at about
17 and 38 MeV \cite{Won10,Won20}.  They were proposed as the parent
particles of the anomalous soft photons observed at CERN
\cite{Chl84,Bot91, Ban93,Bel97,Bel02pi,Bel02,DEL06,DEL08,Per09,DEL10}.
Anomalous neutral bosons with masses at about 17 and about 38 MeV
observed at ATOMKI
\cite{Kra16,Kra19,Nag19,Kra21,Kra22,X17Kra,X1722,Kra23,Sas22},
Dubna\cite{Abr09,Abr12,Abr19,Abr23}, and HUS\cite{Tra24} and were
called the hypothetical X17 and E38 particles respectively.  These
observations hold promising experimental evidence for the QED mesons
and support the possible existence of neutral color-singlet quark
matter at $T$=0, pending further theoretical and experimental
confirmations.

We would like to review here the theoretical and experimental evidence
for QED mesons and color-singlet quark matter.  We wish to study $qqq$
quark matter and to discuss the implications, if the existence of
these objects would be established.

\section{Confined $q\bar q$ states in neutral  color-singlet 
and color-octet 
quark matters at $T$=0}

As we explain in the Introduction, a quark and an antiquark in the
neutral color-singlet quark matter must interact only in the
color-singlet U(1) QED interaction in order to form stable colorless
$[[q\bar q]^1\gamma^1]^1$ complexes at $T$=0.  We note that the
Schwinger confinement mechanism \cite{Sch62,Sch63} stipulates that a
massless fermion and its antiparticle interacting in the Abelian QED
interaction in (1+1)D with a coupling constant $g_\2d^\qedu$ will be
confined as a massive boson with a mass $m^\qedu$,
\begin{eqnarray}
(m^\qedu)^2= \frac{(g_\2d ^\qedu)^2 }{\pi},
\end{eqnarray}
for the  fermion system with a single flavor.

The neutral color-singlet quark matter with two flavors possesses
colorless complexes characterized by $I_z$=$0, I$=$0,1$ involving a
linear combination of $|u \bar u\rangle$ and $|d \bar d\rangle$ states
with unlike-sign charged quarks.  Upon approximating the quarks as
massless, the Schwinger confinement mechanism can be applied to the
color-singlet quark matter to give the masses of the $I_z$=0, $I$=0,1
QED mesons in (1+1)D as \cite{Won10,Won20}
\begin{eqnarray}
(m_{ I}^\qedu)^2 = \frac{(g_\2d ^\qedu)^2  [Q_u ^\qedu+(-1)^I  Q_d^\qedu]^2}{2\pi},
\end{eqnarray}
where $Q_u^\qedu=2/3$ and $Q_d^\qedu=-1/3$  are the electric charge number of the $u$ and $d$ quarks respectively.

A $q\bar q$ QED meson in (1+1)D as an open-string may represent a
physical QED meson in (3+1)D as a flux tube if the $q\bar q$
open-string is the approximate compactification of a physical $q\bar
q$ flux tube in (3+1)D.  The central question is whether the $q\bar q$
QED open-string in (1+1)D is indeed the approximate compactification
of a physical QED flux tube in (3+1)D.  While an $e^+e^-$ system is
confined as an open-string in (1+1)D string, the $e^+e^-$ system is
not confined in (3+1)D because an electron and a positron can be
isolated.  However, for a $q\bar q$ system, a $q\bar q$ open-string
(1+1)D may be an approximate representation of a $q\bar q$ flux tube
in (3+1)D because an isolated quark or antiquark have never been
observed in (3+1)D, and the principle of colorless observable entities
may be at work for a quark and an antiquark in (3+1)D.  Whatever
theoretical debates there may be, the answer to the question will
eventually be settled by experiments.  On that score, the existence of
$q\bar q$ QED mesons in (3+1)D may explain many experimental
anomalies: 1) the soft photon anomaly that whenever hadrons are
produced, soft photons in the form of excess $e^+e^-$ pairs are always
produced, as described in Section 3, and 2) the neutral boson anomaly
at ~17 MeV and ~38 MeV observed at ATOMKI, Dubna, and HUS, as
described in Section 4.  It is therefore useful to propose the
hypothesis that the $q\bar q$ open-string in (1+1)D QED for quarks and
antiquark may be the approximate compactification of a $q\bar q$ QED
flux tube in (3+1)D.

A light quark and a light antiquark interacting in the QCD SU(3)
interaction can be approximated in the quasi-Abelian approximation
\cite{Won10,Won20,Won22,Won22a,Won22c,Won23} as quarks interacting in
the Abelian U(1) interaction with the coupling constants of QCD in
(1+1)D, as in Lund model description of hadrons \cite{And83}.  With
the additional massless quark approximation, we can apply the
Schwinger confinement mechanism to quarks in QCD to infer the mass of
the QCD meson for quarks in (1+1)D as
\begin{eqnarray}
(m^{\qcdu})^2 = \frac{(g_\2d ^\qcdu)^2  }{\pi}.
\end{eqnarray}
For massless quarks with two flavors and the presumed isospin
symmetry, the masses of the $I_z$=0, $I$=0,1 QED and QCD mesons in
(1+1)D are \cite{Won10,Won20}
\begin{eqnarray}
(m_I^{\lambda} )^2\! =\! \frac{(g_\2d ^\lambda)^2  [Q_u^\lambda +(-1)^I  Q_d^\lambda]^2}{2\pi}, 
~\lambda\!=\!\!
\begin{cases}
0  {\rm ~for~QED,}\\
1  {\rm ~for~QCD},\\
\end{cases}\!\!\!\!
\label{qedm}
\end{eqnarray}
where $Q_q^0$  and $Q_q^1$ are  respectively
the electric and color charge numbers of the $q$ quark. 

Previously, when we compactify a system of a flux tube in (3+1)D with
cylindrical symmetry into an open string in (1+1)D, we find that the
longitudinal equation in (1+1)D contains a coupling constant $g_\2d$
that encodes the information of the flux tube transverse radius $R_T$
and the coupling constant $g_\4d$ in (3+1)D as
\cite{Won09,Won20,Won22,Won22c,Won23,Won24}
\begin{eqnarray}
(g_{\2d})^2=\frac{1}{\pi
    R_T^2}(g_{\4d})^2=\frac{4\alpha_{\4d}}{R_T^2}.
\label{12}
\end{eqnarray}
Thus in (3+1)D, the masses  of  QCD and  QED  $I_z$=0, $I$=0,1  mesons  are   
\begin{eqnarray}
(m_{ I}^\lambda)^2 = \frac{4\alpha _\4d^\lambda}{\pi R_T^2}\, \frac{[Q_u^\lambda  +(-1)^I  Q_d^\lambda ]^2}{2}.
\label{qcdm}
\end{eqnarray}
Due to a lack of quantitative information, we shall start with the
working hypothesis that the flux tube radius $R_T$ may be an intrinsic
property of the quarks so that it is the same for QED and QCD for
which $R_T \sim 0.4$ fm \cite{Cos17}.  Such a hypothesis is also
justified {\it a posteriori} for yielding anomalous soft photons and
QED mesons with masses in the appropriate energy region.  In this QED
case with $Q_u^\qedu$=$ 2/3$, $Q_d^\qedu =-1/3$, and
$g_\4d^\qedu=1/137$, we get from Eq.\ ({\ref{qedm}})
$m_{I=0}^\qedu=11.4$ MeV, and $m_{I=1}^\qedu=33.6$ MeV
\cite{Won10,Won20}.

For QCD mesons, the mass of $\pi^0$ is zero from the Schwinger
confinement mechanism and the $\pi^0$ mass receives contributions only
from the quark condensate.  We know from the Gell-Mann-Oakes-Renner
relation \cite{Gel68} that the QCD quark condensate, $\langle \bar
\psi \psi \rangle_{{}_\qcdd}$, contributes to the square of the pion
quark mass.  For QED mesons, there should be likewise a QED quark
condensate, $\langle \bar \psi \psi \rangle_{{}_\qedd}$, contribution
to the square of the QED meson mass.  As the quark condensate arises
from the interaction between the quark and the antiquark \cite{Pes95}
and the interaction between a quark and an antiquark depends on the
strength of the coupling constant $\alpha_\4d$, we therefore expect
the QCD and QED quark condensate contributions to the composite mass
square $\Delta m^2$ to be proportional to the square of their
respective coupling constants $(g_\4d^\lambda)^2$, or $\Delta
m_\qedd^2/\Delta m_\qcdd^2$$\sim$$\langle \bar \psi \psi
\rangle_{{}_\qedd}$$/$$\langle \bar \psi \psi
\rangle_{{}_\qcdd}$\!\!$\sim$\,$\alpha_\4d^\qedu$$/$$\alpha_\4d^\qcdu$.
The mass formula for QED mesons with two flavors and the quark
condensate can be estimated to be \cite{Won20}
\begin{eqnarray}
(m_{ I}^\qedu)^2 = \frac{4\alpha _\4d^\qedu}{\pi R_T^2}\, \frac{[Q_u^\qedu +(-1)^I  Q_d^\qedu]^2}{2}  + m_\pi^2 \frac{\alpha_\4d^\qedu}{\alpha_\4d^\qcdu} ,  
\end{eqnarray}
 which gives $m_{I=0}^\qedu$=$17.9$MeV, $m_{I=1}^\qedu$=$36.4$MeV.  

For QCD mesons, it is necessary to generalize the mass formula to
three flavors with the inclusion of the strange quark to give
\begin{eqnarray}
m_i^2=\frac{4\alpha_\4d^\qcdu}{ \pi R_T^2} (\sum_{f=1}^{3} D_{if}Q_f^\qcdu)^2  +
m_\pi^2 \sum_{f=1}^{3} \frac{m_f}{m_{ud}} (D_{if})^2,
\label{qcd}
\end{eqnarray}
where 
$m_{ud}=(m_u+m_d)/2$, and 
the physical meson is $\Phi_i$=$\sum_{f=1}^3 D_{if} \phi_f$, $\phi_1$=$|u\bar u\rangle,  \phi_2$=$|d\bar d\rangle, \phi_3$=$|s \bar s\rangle$.  
With $Q_u^\qcdu$=$Q_d^\qcdu $=$1$, and   $g_\4d^\qcdu=0.68$ MeV,
the above mass formula gives 
$m_\eta=498.4$ MeV, and $m_{\eta'}$=948.2 MeV, in approximate agreement with experimental values \cite{Won20}.

To search for the signature of the neutral color-singlet quark matter
at $T$=0, our task is to find the QED meson states with these masses
in hadron collisions and $e^+$-$e^-$ annihilations in which $q\bar q$
pairs may be produced.  We expect that these bosons will decay into
$e^+$-$e^-$ or $\gamma\gamma$ pairs
\cite{Won20,Won22,Won22c,Won23,Won24} .

\section{Observations of the anomalous soft photons at CERN and the DELPHI anomaly}

Experimentally, there were numerous observations of excess $e^+$$e^-$
pairs at CERN, labeled as ``anomalous soft photons'', whenever hadrons
are produced in high-energy $K^+ p$ \cite{Chl84,Bot91}, $\pi^+ p$
\cite{Bot91}, $\pi^- p$ \cite{Ban93,Bel97,Bel02pi}, $pp$ collisions
\cite{Bel02}, and $e^+$$e^-$ annihilations
\cite{DEL06,DEL08,Per09,DEL10}.  Specifically in the DELPHI exclusive
measurements in the decay of $Z^0$ in $e^+ e^-$ annihilations, the
excess $e^+$$e^-$ pairs were observed to be proportionally produced
when hadrons (mostly mesons) were produced \cite{Per09,DEL10} (see
Fig.\ \ref{DELPHI}(b)), and they were not produced when hadrons were
not produced \cite{DEL08}.  The transverse momenta of the excess
$e^+$$e^-$ pairs lied in the range of a few MeV/c to many tens of
MeV/c, corresponding to a mass scale of the anomalous soft photons in
the range from a few MeV to many tens of MeV.

Owing to the simultaneous and correlated production alongside with
hadrons, a parent particle of the anomalous soft photons is likely to
contain elements of the hadron sector, such as a light quark and a
light antiquark.  Relative to the QCD interaction, the Schwinger
mechanism for massless quarks in (1+1)D QED interaction will bring the
quantized mass of a $q\bar q$ pair in Eq.\ (1) to the mass range of
the anomalous soft photons of many tens of MeV.  It was therefore
proposed in \cite{Won10} that a quark and an antiquark in a $q\bar q$
system interacting in the QED interaction might lead to new open
string bound states (QED-meson states) with a mass of many tens of
MeV. These QED mesons might be produced simultaneously with the QCD
mesons in the string fragmentation process in high-energy collisions
\cite{Chl84,Bot91,Ban93,Bel97,Bel02pi,Bel02,DEL06,DEL08,Per09,DEL10},
and the excess $e^+e^-$ pairs might arise from the decays of these QED
mesons.  Theoretical calculations based on such an open-string model
gave a good description of the $p_T$ distribution and the mass region
of the anomalous soft photons \cite{Won10,Won20}

\begin{figure} [h]
\centering
\vspace*{-0.0cm}
\hspace*{-0.5cm}
\includegraphics[scale=0.82]{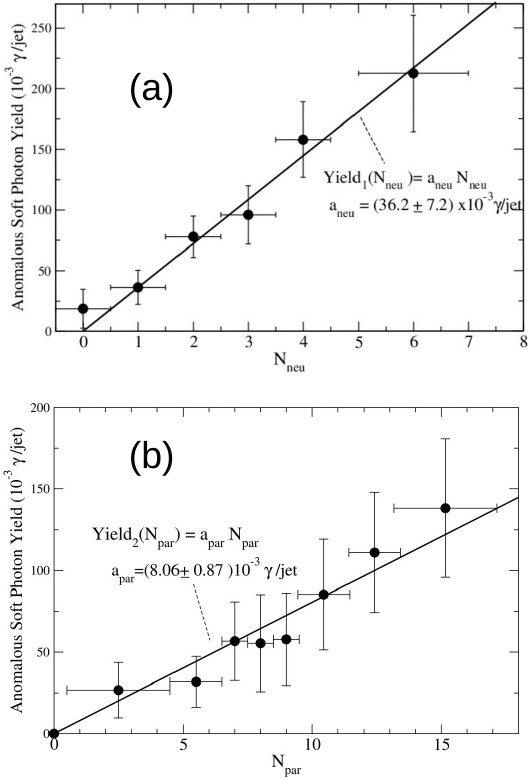}
\vspace*{-0.2cm}
\caption{ The anomalous soft photon yield in the $e^+e^-$ annihilation
  at $Z^0$ energy in DELPHI measurements at CERN \cite{DEL10}.  In
  Fig.\ 2(a), (Yield$_1$) is the yield as a function of the number of
  neutral particles, $N_{\rm neu}$, inclusive of different number of
  charged particles, $N_{\rm ch}$.  In Fig.\ 2(b), (Yield$_2$) is the
  yield as a function of $N_{\rm par}$=$N_{\rm neu}$+$N_{\rm ch}$, the
  total number of neutral and charged particles. }
  \label{DELPHI}
\end{figure}

There is a perplexing DELPHI anomalous soft photon anomaly that
warrants special attention.  In the $e^+e^-$ annihilation at the $Z^0$
energy, the production of the anomalous soft photons was found to be
more probable in conjunction with the production of neutral hadrons
than with charged hadrons, in contrast to the common intuitive notion
that the electromagnetic radiation (such as photons) are associated
more likely with charged particle production than with neutral
particle production.  The yield $y_{\rm asp}$ of the anomalous soft
photons is found to relate to the number of produced charged hadrons
$N_{\rm ch}$ and the number of produced neutral hadrons $N_{\rm neu}$
by \cite{DEL10}
\begin{eqnarray}
y_{\rm asp} = a_1 N_{\rm ch} + a_2 N_{\rm neu},
\label{eq8}
\end{eqnarray}
where $a_1=dy_{\rm asp}/dN_{\rm ch} =(6.9\pm 1.8\pm 1.8)\times
10^{-3}$ $\gamma$/jet and $a_2=dy_{\rm asp}/dN_{\rm neu} =(37.7 \pm
3.0\pm 3.6)\times 10^{-3}$ $\gamma$/jet, and $dy_{\rm asp}/dN_{\rm
  neu} \gg dy_{\rm asp}/dN_{\rm ch}$ \cite{DEL10}.  Similarly, Figure
2(a) gives $dy_{\rm asp}/dN_{\rm neu}=(36.2 \pm 7.2)\times 10^{-3}$
$\gamma$/jet and Fig.\ 2(b) gives $dy_{\rm asp}/dN_{\rm par}=(10.0 \pm
0.87)\times 10^{-3}$ $\gamma$/jet.  The Schwinger pair production
mechanism \cite{Won10} and dipole radiation \cite{DEL10} were
suggested as possible explanations of such an unusual DELPHI anomaly.
However, in view of the different properties of the neutral and
charged color-singlet quark matters as presented in Eq.\ (\ref{eq8})
and the Introcution, a better alternative explanation may be more
appropriate.

We envisage that the dynamics of the quark matter at $T$=0 can be
described as the space-time variation of the quark current and the
gauge field at each space-time arena as in a quantum fluid.  The
dynamics of the mesons is characterized by stable, localized,
collective, and periodic space-time variations of the quark current
$j^\mu(x,t)$ and the gauge field $A^\mu(x,t)$
\cite{Won10,Won20,Won22c,Won23,Won24}.  The QCD mesons $\pi^0$,
$\eta^0$,... are excitation quanta of the color-octet quark fluid at
$T$=0.  The neutral color-singlet quark matter at $T$=0 distinguishes
itself by having localized stable and confined excitation quanta of
the fluid in the form of the QED $q \bar q$ mesons whereas the charged
color-singlet quark matter does not possess stable QED excitation
quanta at $T$=0.

Hence, in situations in which the quark fluid is excited by an
external disturbance such as in the production of a neutral or a
charged $q\bar q$ pair in string fragmentation process, both the
neutral color-octet and neutral color-singlet quark matters can be
excited to lead to the production of the QCD and/or QED $q\bar q$
mesons when a neutral $q\bar q$ pair is produced.  When a charged
$q\bar q$ pair is produced in the string fragmentation process, the
charged color-octet quark matter excitation can lead to the production
of stable QCD $q\bar q$ mesons but the charged color-singlet quark
matter does not possess nor produce stable excitation quanta at $T$=0.
However, at a higher $T\ne 0$ temperature, the charge color-singlet
quark matter may contain a small amount of neutral QED meson gas in a
thermal environment.  Neutral QED mesons in thermal equilibrium
amounts may be produced in conjunction with the production of charged
QCD mesons at $T\ne 0$ .  Consequently, neutral QED mesons, the
precursors of the anomalous soft photons, are more likely to be
produced in association with neutral QCD mesons than with charged QCD
mesons.  It will be of great interest to study further whether the
anomalous soft photon production preference for neutral QCD meson in
the DELPHI anomaly may be used to differentiate the properties of the
neutral and charged color-singlet quark matters at different
temperatures.

\section{ Observations of anomalous bosons with masses about 17 and 38 MeV}

\subsection{ATOMKI observations of X17}

Since 2016, the ATOMKI Collaboration have been observing the
occurrence of a neutral boson, the hypothetical X17, with a mass of
about 17 MeV, by studying the $e^+ e^-$ spectrum in the de-excitation
of the excited alpha-conjugate nuclei $^4$He$^*$, $^8$Be$^*$,
$^{12}$C$^*$ at various energies in low energy proton-fusion
experiments \cite{Kra16,Kra21,Kra22,Kra23}.  A summary and update of
the ATOMKI results were presented \cite{X17Kra,Kra23} and the
confirmation of the ATOMKI data for the $p+{}^7$Be reaction by Hanoi
University of Science (HUS) was reported \cite{Tra24}.  The signature
for the X17 particle consists of a resonance structure in the
invariant mass of the emitted $e^+e^-$ pair.  Such a signature is a
unique identification of a particle.  Because $e^+e^-$ pairs are also
produced in the internal pair conversion of the photon from the
radiative decay of the compound nucleus, it is necessary to subtract
contributions from such a process and from random and cosmic-ray
backgrounds.

The ATOMKI experiments consist of low-energy proton fusion of the
nucleus $A$ (the proton) and nucleus $B$ (the target nucleus) to
produce an $\alpha$-conjugate nucleus $C^*$ with the subsequent
$e^+e^-$ de-excitation of the $C^*$ nucleus to the ground state
$C_{\rm gs}$ or an excited state $C_{\rm f}$ in the reaction
\begin{eqnarray}
 A + B \to C^* \xrightarrow{\gamma}  C_{\rm gs} ~{\rm or ~} C_{\rm f} ,
\end{eqnarray}
which is  followed~by
\begin{eqnarray}
&&\hspace*{-1.2cm}\gamma\xrightarrow{{\rm internal ~conversion}}  e^+ + e^- ,
\nonumber\\
\text{and if}&&\hspace*{-0.2cm} C^*\!\!\xrightarrow{\gamma}C_{\rm f}, ~{\rm then~} C_{\rm f} \to C_{\rm gs} + \gamma_{\rm f} .
\end{eqnarray}
The internal pair conversion process has a characteristic
$dN/d\Omega_{\theta_{e^+e^-}}$ distribution of the opening angle
between $e^+$ and $e^-$ that depends on the multi-polarity of the
$C^*$$\xrightarrow{\gamma} C_{\rm gs}$ or $C_{\rm f}$ transition
\cite{Ros50}.  It is a relatively smooth
$dN/d\Omega_{\theta_{e^+e^-}}$ distribution in the CM system in which
$C^*$ is at rest.

In the presence of such an $e^+e^-$ internal pair conversion
background, the ATOMKI Collaboration searches for an unknown neutral
boson X that may be emitted by $C^*$ in its de-excitation,
\begin{eqnarray}
A + B \to C^* \xrightarrow{X}  C_{\rm gs}  ~{\rm or~} C_{\rm f}  , 
\end{eqnarray}
which is followed by
\begin{eqnarray}
&&\hspace*{-1.2cm}
X \to e^+ + e^- ,
\nonumber\\
\text{and if}&&\hspace*{-0.2cm} C^*\!\!\xrightarrow{X}C_{\rm f}, ~{\rm then~} C_{\rm f} \to C_{\rm gs} + \gamma_{\rm f} .
\end{eqnarray}
The decay of the $X$ particle into $e^+$ and $e^-$ give an $e^+e^-$
excess above the internal pair conversion $e^+e^-$ background, as the
signal for the neutral X boson.

 In the $C^*$ center-of-mass system, the decay of an unknown X
 particle with a mass $m_X$ into $e^+$ and $e^-$ will lead to an
 $e^+$-$e^-$ energy sum $E_{e^+e^-}$ distribution,
\begin{eqnarray}
dN/dE_{e^+e^-}=\delta(E_{e^+e^-}-K-m_X),
\end{eqnarray}
 and  an $e^+$-$e^-$
opening angle $\theta_{e^+e^-}$ distribution  
\begin{eqnarray}
\frac{dP}{d\Omega_{\theta_{e^+e^-}}}
&=& \frac{1}{4\pi} \frac  {1} {\gamma^2 \beta (1-\cos\theta_{e^+e^-})^{3/2}}
\nonumber\\
& & \times{\frac{1}{
\sqrt{{-1+2\beta^2- \cos\theta_{e^+e^-}}} } },
\label{dpm}
\end{eqnarray}
as obtained by McDonald \cite{Mc76} in the similar case for $\pi^0$
decay into two photons.  The minimum opening angle
$\theta_{e^+e^-}$(min) in the observer system is given by
\begin{eqnarray}
\cos [\theta_{e^+e^-}({\rm min})] =-1 + 2\beta^2,
\label{min}
\end{eqnarray}
as was obtained earlier by McDonald in \cite{Mc76}, Barducci and Toni
in \cite{Bar23}.

\begin{figure} [h]
\centering
\includegraphics[scale=0.40]{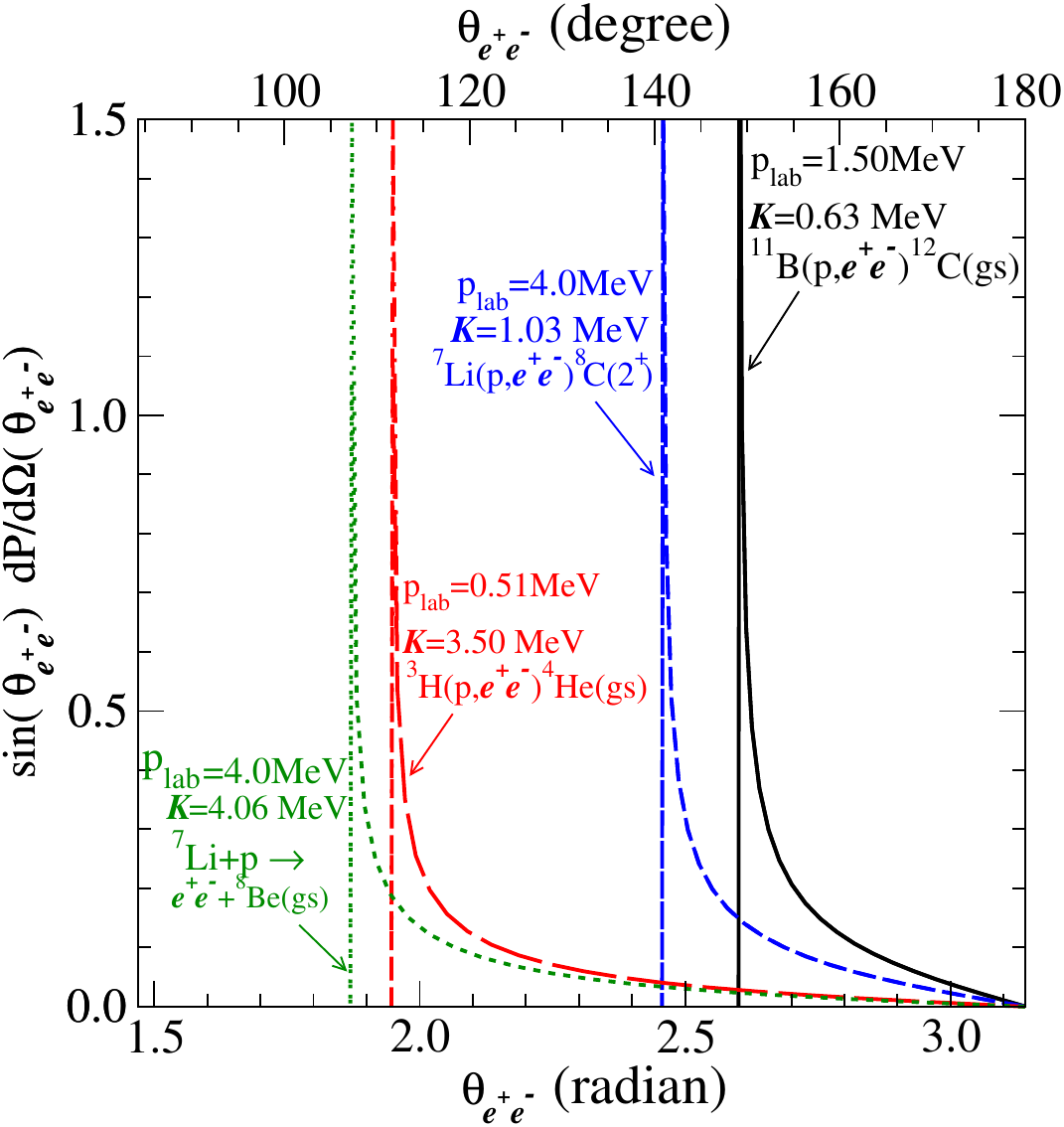}
\caption{ The $e^+e^-$ opening angle distribution, $\sin(
  \theta_{e^+e^- })dP /d\Omega_{\theta_{e^+e^- }}$, for different
  proton $E_p^{\rm lab}$, different kinetic energies $K$ of the X17
  boson, and different reaction initial participants and final compound
  nucleus states, for $m_{X17}$=16.7 MeV.  }
  \label{angdis}
\end{figure}

The quantity $K$ in Eq.\ (14) is the sum of the kinetic energy of the
X boson and the kinetic energy of its emission partner $C_{\rm f}$ in
the CM system given explicitly by
\begin{eqnarray}
&&K=  E_x  -[M(C_{\rm f})-M(C_{\rm gs})] -m_X,
\\
&&E_x=  \frac{AB}{A+B} E_A^{\rm lab}+  Q_{\rm gs} ,
\\
&&Q_{\rm gs}=  M(A) + M(B)-M(C_{\rm gs}),
\label{last}
\end{eqnarray}
where $E_x$ is the excitation energy of the compound nucleus $C^*$
relative to its ground state $C_{\rm gs}$, 
 $M(A)$, $M(B)$, $M(C)$, and $M(C_{\rm f})$ are the masses  of $A$, $B$, $C$, and $C_{\rm f}$ 
respectively, $m_X$ is the
mass of the X boson, and $Q_{\rm gs}$ is the $Q$ value for $AB\to C_{\rm gs}$.   As $ M(C_{\rm f}) \gg m_X$,  $K$ is
essentially the kinetic energy of X in the $C^*$ CM system.
The quantity $\beta^2$
is related to $K$ by
\begin{eqnarray}
\beta^2&&
=1-\frac{1}{(1+K/m_X)^2}
\nonumber\\
&&
= 
 \left (\frac{2 K}{m_X} +
 \frac{K^2}{m_X^2}
 \right ) \biggr / 
 \left  ( 1 + \frac{2 K}{m_X} 
 +
 \frac{K^2}{m_X^2} \right ).
\label{eq6}
\end{eqnarray}

 The opening angle distribution for a few cases in the ATOMKI
 experiment for the emission of the $X$ particle are shown in Fig.\ 3.

At the minimum opening angle $\theta_{e^+e^-}$(min), the denominator
on the right hand side of Eq.\ (\ref{dpm}) goes to zero, and the
opening angle distribution ${dP}/{d\Omega _{\theta_{e^+e-}}} $
diverges to infinity and gives rise to the so-called ``Jacobian
peak"\footnote{The author is indebted to Dr.\ T.\ Awes for pointing
out the "Jacobian Peak" name for such a peak.}  in Fig.\ \ref{angdis}.

\begin{figure}[h]
\centering
\includegraphics[scale=0.50]{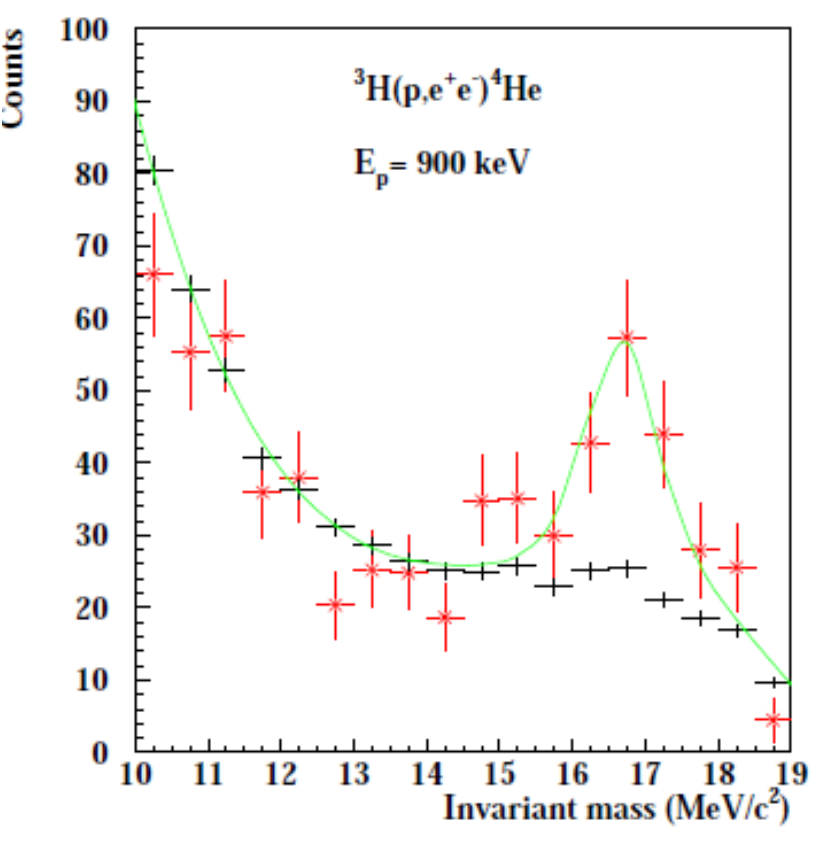} 
\caption{ The invariant mass distribution of the emitted $e^+$ and
  $e^-$ in the de-excitation of the compound nucleus $^4$He$^*$ state
  at 20.49 MeV in the $^3$H($p,e^+e^-){}^4$He$_{\rm gs}$ reaction at
  $E_p^{\rm lab}=$0.9 MeV as given in \cite{Kra19}.  Red data points are the data
  in the signal region, $19.5 < E_{e^+e^-}<$ 22.0
  MeV, and black points are data in the background region, $5 <
  E_{e^+e^-}<$ 19.0 MeV .  }
\label{thetaK}
\end{figure}

For an optimal detection of the X17 signals, the ATOMKI Collaboration
found it necessary to focus on certain regions of the phases space
with strong signals so as to enhance the observation probability.  For
example, in the collision of $p$ with $^3$H at 0.9 MeV energy, ATOMKI
Collaboration found that the correlation angle $\theta_{e^+e^-}$ of
120$^o$ was optimal for a large signal (see Fig.\ 3). At such an
angle, the energy sum of the $e^+$ and $e^-$, $E_{e^+e^-}({\rm
  sum})=E_{e^+}+E_{e^-}$, showed a peak structure at around $20.6$ MeV
as shown in Fig.\ 1 of \cite{Kra21}.  Two spectra were constructed for
the energy sum $E_{e^+e^-}({\rm sum})$, one at
$\theta_{e^+e^-}$=120$^o$ and another at 60$^o$ where no X17 signal
was expected.  The energy sum spectrum in the lower panel of Fig.\ 1
of \cite{Kra21} was obtained by subtracting the latter from the
former, after proper normalization.  In the signal region of $19.5 \le
E_{e^+e^-}\le 22.0$ MeV and the background in $5 \le E_{e+e^-}<19$
MeV, the invariant mass spectrum of the emitted $e^+$ and $e^-$ showed
a resonance structure at $\sim$ 17 MeV as shown in
Fig.\ {\ref{thetaK}}.

In the ATOMKI X17 emission model, the nature of the X17 and the
coupling between the emitting excited alpha conjugate nucleus and the
emitted X17 particle are left unspecified.  Among many possibilities,
the X17 was suggested as a possible carrier of a fifth force
\cite{Fen16,Fen17,Fen20} and has generated a great deal of interest
\cite{X1722}.  It could also be the isoscalar QED meson, which has a
predicted mass of $\sim$17 MeV and can decay into $e^+$ and $e^-$
\cite{Won10,Won20}.  The nature of the particle can be determined only
by future experiments.

Among many proposed models for the X17 particle, there is one that
involves QED explanation presented by Varr\' o \cite{Var24} who
suggests that the X17 particle may be a dressed radiation excitations
of photons that gain a non-zero rest mass through their own
$A^\mu$-field self-interaction and their interaction with a proton
while the E38 may be a similar dressed radiation excitations of
photons that gain a non-zero rest mass through their interaction with
a neutron.  It will be of interest to study in future work whether the
dressed self-interacting photon of Ref. \cite{Var24} may be related to
the $A^\mu$$\to$$j^\mu$$\to$$A^\mu$$\to$$j^\mu$... loop (the AjA
nonlinearity) that will lead to photon $A^\mu$ self-interaction and
QED mesons involving massless quarks of the Schwinger confinement
mechanism in \cite{Won93, Won10,Won20,Won22,Won22c}.
 
 \subsection{  $\theta_{e^+e^-}$(min)  systematics}

The successful analyses of the total $e^+ e^-$ opening angle
distribution is subject to the uncertainties of the $e^+ e^-$
background contributions from internal pair conversion
\cite{Zha17,Hay22}.  It is useful to make a supplementary test on the
X17 signals from another robust perspective.  Specifically, because of
the big difference in the shapes of the smooth opening angle
distribution from the internal pair conversion background and the
sharp Jacobian peak opening angle distribution from the X17 decay
products, the onset of the additional X17 decay product contributions
can be obtained with a lower degree of uncertainty.  As shown in
Fig.\ \ref{angdis}, the occurrence of the X17 particle is signaled by
a sudden rise of the opening angle distribution as a function of the
opening angle because of the occurrence of the Jacobian peak, and
correspondingly a sudden change of the slope of the total opening
angle distribution at $\theta_{e^+e^-}$(min).

\begin{figure}[h]
\centering
\includegraphics[scale=0.33]{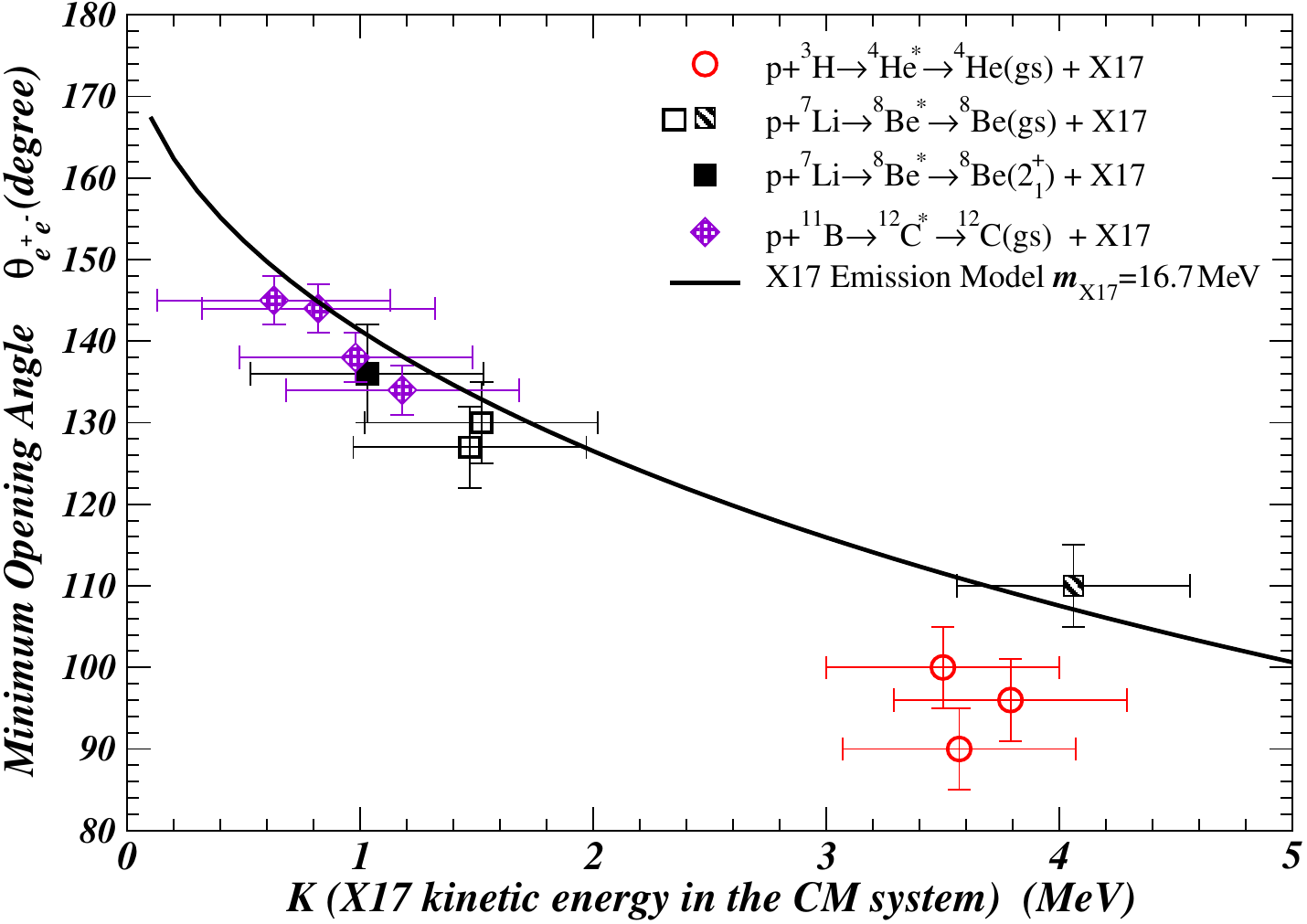}
\caption{ Comparison of the experimental data of the minimum opening
  angle $\theta_{e^+e^-}$(min) as a function of the X17 kinetic energy
  $K$ from ATOMKI \cite{Kra16,Kra21,Kra22,Kra23} and HUS \cite{Tra24}
  for different collision energies, targets, and final states.  The
  X17 emission model envisages the fusion of the incident proton $p$
  with the target nucleus $B$ forming a compound nucleus $C^*$, which
  subsequently de-excites to the final state $C_{\rm f}$ with the
  simultaneous emission of the X17 particle.  Subsequent decay of the
  X17 particle into $e^+$ and $e^-$ then gives the angle
  $\theta_{e^+e^-}$ between $e^+$ and $e^-$.  The curves give the
  theoretical predictions of $\theta_{e^+e^-}$(min) as a function of
  the X17 kinetic energy $K$.  }
 \label{minthK}
\end{figure}

From each of the experimental ATOMKI reactions in
\cite{Kra16,Kra19,Nag19,Kra21,Kra22,X17Kra,X1722,Kra23,Sas22}, and the
HUS reaction in \cite{Tra24}, we extract the $\theta_{e^+e^-}$(min)
value to be the midpoint between a sudden jump in the first derivative
of the angular distribution of $\theta_{e^+e^-}$ at the onset of the
anomaly.
   
We compare in Fig.\ {\ref{minthK}} the experimental
$\theta_{e^+e^-}$(min) data as a function of $K$ with the theoretical
predictions of the X17 emission model, obtained from Eqs.\ (\ref{min})
and (\ref{eq6}).  All final nucleus states $C_{\rm f}$ are implicitly
ground states, except specified explicitly for the case of the
$^7$Li$(p,e^+e^-)^8$Be$(2^+)$ reaction at $E_p^{\rm lab}$=4 MeV.  One
observes that there is a reasonable agreement between the theoretical
curve and the data point in Fig.\ \ref{minthK} for all cases of
collision energies, initial colliding nuclei, and final compound
nucleus states, indicating the approximate validity of the ATOMKI X17
emission model \cite{Kra16}.  Previous analyses on the X17 decay by
Feng $et~al.$ \cite{Fen20} and by Barducci and Toni \cite{Bar23} using
earlier experimental data also support the validity of the ATOMKI X17
emission model \cite{Kra16}.

Recently, ATOMKI reported the observation of the emission of the X17
particle in the de-excitation of the compound nucleus ${}^8$Be$^*$ to
the excited ${}^8$Be($2^+$,3.03MeV) state \cite{Kra23}.  We can test
the extended ATOMKI X17 emission model \cite{Kra23} which proposes the
emission of the X17 particle not only in the de-excitation of the
produced compound nucleus to the ground state, but also to an excited
state of the compound nucleus \cite{Kra16}.  In Fig.\ {\ref{minthK}}
the experimental minimum opening angle $\theta_{e^+e^-}$(min) for this
de-excitation to the excited $2^+$ state is shown as the solid square
data point.  This data point follows the same systematics as other
reaction data points for de-excitation down to the ground states.  The
ATOMKI X17 emission model is therefore shown to be valid also for the
de-excitation of the compound nucleus $C^*$ to an excited state
$C_{\rm f}$ of the compound nucleus.

We would like to remark that although the systematics of the
experimental minimum opening angle $\theta_{e^+e^-}$(min) are in
general agreement with the X17 emission model, there is however a
notable small difference between the data points for $^4$He and those
following the systematics for ${}^8$Be and ${}^{12}$C in
Fig.\ \ref{minthK}.  It is well known that ${}^4$He is a spherical
nucleus while ${}^8$Be and ${}^{12}$C are strongly deformed prolate
and oblate nuclei, respectively. It will be of great interest to
investigate theoretically whether the minor difference in
$\theta_{e^+e^-}$(min) systematics may arise from the final state
interaction of very deformed $^8$Be and $^{12}$C nuclei in modifying
the opening angles between $e^+$ and $e^-$.

\subsection{
  The HUS observation of X17 with a new $e^+e^-$ spectrometer}

The ATOMKI $e^+e^-$ spectrometer had five arms in 2016 \cite{Kra16}.
The spectrometer was subsequently improved to acquire an additional
sixth arm.  Taking advantages of the occurrence of the optimal opening
angles in the decay of the hypothetical X17 particle, Tran $et~al.$
\cite{Tra24} at Hanoi University of Science, in collaboration with
ATOMKI experimentalists, built a new two-arm $e^+e^-$ spectrometer to
check the ATOMKI $^8$Be anomaly.  The HUS spectrometer consists of
only two arms making an angle of 140$^o$ with respect to each other.
It bases its design on the theoretical and experimental estimate that
when the proton beam is at $E_{\rm lab}=$1.04 MeV, the opening angle
$\theta_{e^+e^-}$ at 140$^o$ will be optimal for the detection of both
$e^+$ and $e^-$.

\begin{figure}[h]
\centering
\includegraphics[scale=0.40]{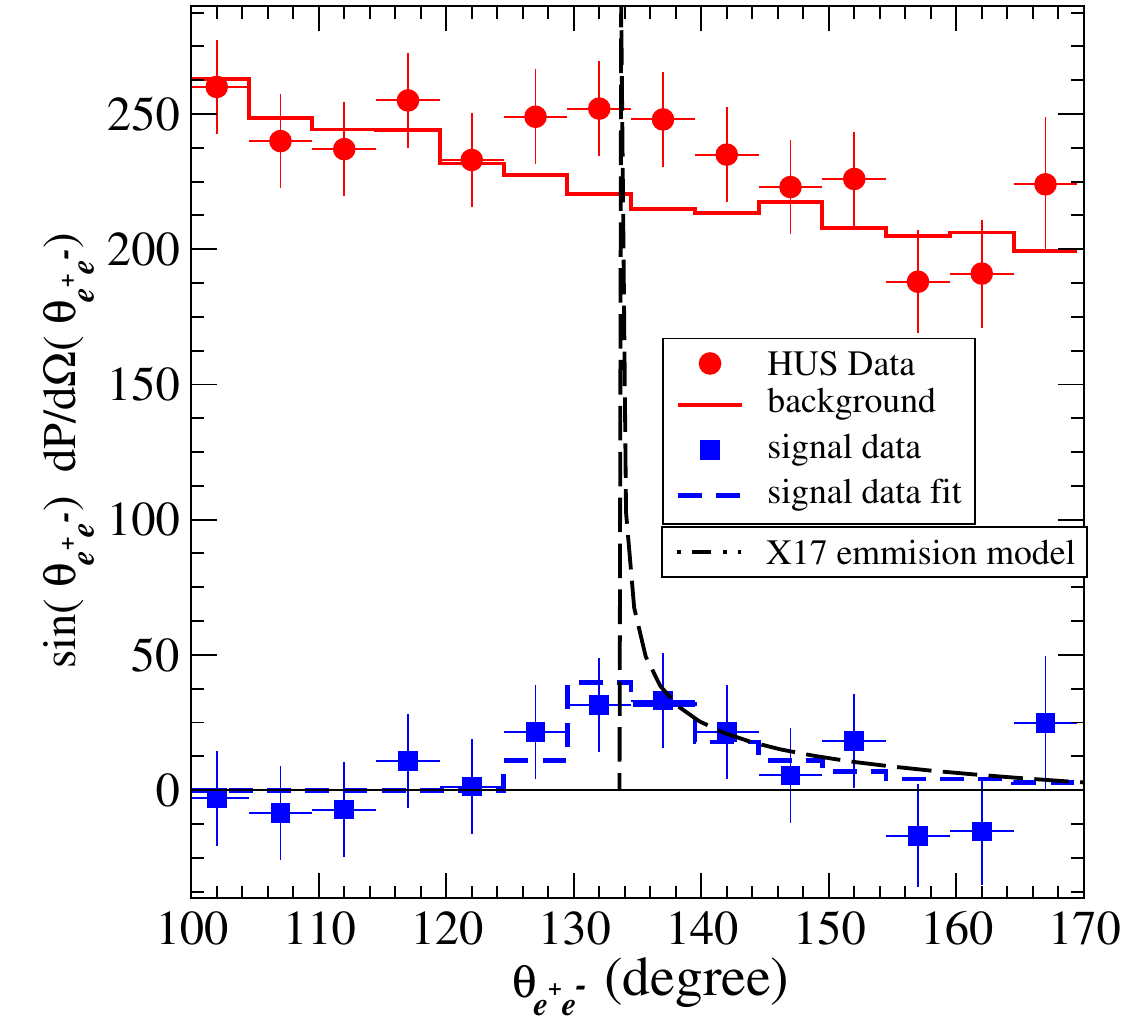} 
\caption{ Experimental $\sin(\theta_{e^+e^-})
  dN/d\Omega_{\theta_{e^+e^-}}$ data from Hanoi University of Science
  (HUS) in $^7$Li$(p,e^+e^-)^8$Be$_{\rm gs}$ reactions at $E_p=1.04$ MeV as given
  in \cite{Tra24}.  The solid circles give the opening angle
  distribution data for $e^+$ and $e^-$ and the solid curve gives the
  background opening angle distribution.  The difference of
  data-background is give by the solid square data points and the
  dashed curve is the HUS fit to the signal data \cite{Tra24}.  The
  dash-dot curve with a Jacobian peak is the theoretical opening angle distribution from
  the X17 emission model calculated with Eq.\ (\ref{dpm}).}
\label{HUS}
\end{figure}

In the HUS measurement, the experimental opening angle distribution
data points are shown as solid circles in Fig.\ {\ref{HUS}}.  The
background distribution arising from the sum of E1 and M1 internal
pair conversion transitions is shown as the solid curve in
Fig.\ {\ref{HUS}}.  There is an $e^+e^-$ excess of event counts as the
experimental opening angle distribution exhibits a deviation above the
solid curve of the internal pair conversion background.  The
difference of the experimental $e^+e^-$ data counts and the background
distributions are the X17 signals which are shown as solid square data
point in Fig.\ {\ref{HUS}}.  The experimental X17 signal can be
described adequately by the opening angle distribution arising from
the decay of the X17 particle with a mass of 16.7 MeV, when the
uncertainties in the event number fluctuations and in the opening
angle measurements are taking into account, as presented by HUS shown
as the dashed curve in Fig.\ {\ref{HUS}} \cite{Tra24}.

The successful observation of the X17 signal with an
optimally-designed $e^+e^-$ spectrometer with only two arms (in lieu
of ATOMKI's five or six arms) by Tran $et~al.$ \cite{Tra24} is a
notable indication of the validity of the X17 emission model and the
possible existence of the X17 particle.  The statistics of the
observed opening angle distribution could be improved with a longer
experimental running period to provide a greater support for the
existence of the X17 particle.

There are other searches for the X17 particle using different methods
\cite{X1722}.  Recently, the PADME Collaboration searched for the X17
particle by scattering $e^+$ and $e^-$ at around the resonance energy,
using a positron beam to collide with electrons in a diamond target.
They obtained a resonance signal at the expected energy with a
statistical measure of about 2$\sigma$ magnitude \cite{PAD25}, which
is not yet significant enough for a definitive confirmation.  A
measurement with a greater statistical significance is needed.
Another MEG II experiment, originally designed to search for the
$\mu^+ \to e^+ \gamma$ decay, was adapted to investigate the X17 by
studying the $p(^7{\rm Li},e^+e^-)^8$Be reaction, with a proton beam
at an energy at 1.080 MeV colliding with the $^7$Li target, resulting
in the excitation of two different resonances \cite{MEG25}.  The
sensitivity of the measurement has not yet reach the level to observe
a significant signal, and limits on the branching ratios of the two
resonances to X17 were set.

\subsection{ The Dubna observation of diphoton decays of X17 and E38}

Abraamyan and collaborators at Dubna have been investigating the
two-photon decay of particles to study the resonance structure of the
lightest hadrons and $q\bar q$ states, using $d$ and $p$ beams of a
few GeV with fixed internal C and Cu targets at the JINR Nuclotron
\cite{Abr09,Abr12,Abr19,Abr23}.  Their PHONTON2 detector consists of
two arms placed at 26 and 28 degrees from the beam direction, with
each arm equipped with 32 lead-glass photon detectors.  The photon
detectors measure the energies and the emission angles of the photons,
from which the invariant masses of the photon pairs can be measured.
By selecting photon pairs from the same arm with small opening angles,
it is possible to study neutral bosons with small invariant masses
such as those below the pion mass gap $m_\pi$.  Upon the suggestion of
van Beveren and Rupp \cite{Bev11}, they search for a neutral boson
resonance with a mass of 38 MeV, they reported earlier the observation
of a resonance structure at a mass of $\sim$38 MeV \cite{Abr12,Abr19}.
In a recent analysis in the diphoton spectrum extended down to the
lower invariant mass region, the Dubna Collaboration reported the
observation of resonance-like structures both at $\sim$17 and $\sim$
38 MeV in the same experimental set-up in which the decay of the
$\pi^0$ particle into two photons was also observed, in support of
earlier ATOMKI observation of the hypothetical X17 particle
\cite{Abr23}.

\begin{figure} [h]
\centering
\vspace*{-0.3cm}
\hspace*{-0.2cm}
\includegraphics[scale=0.57]{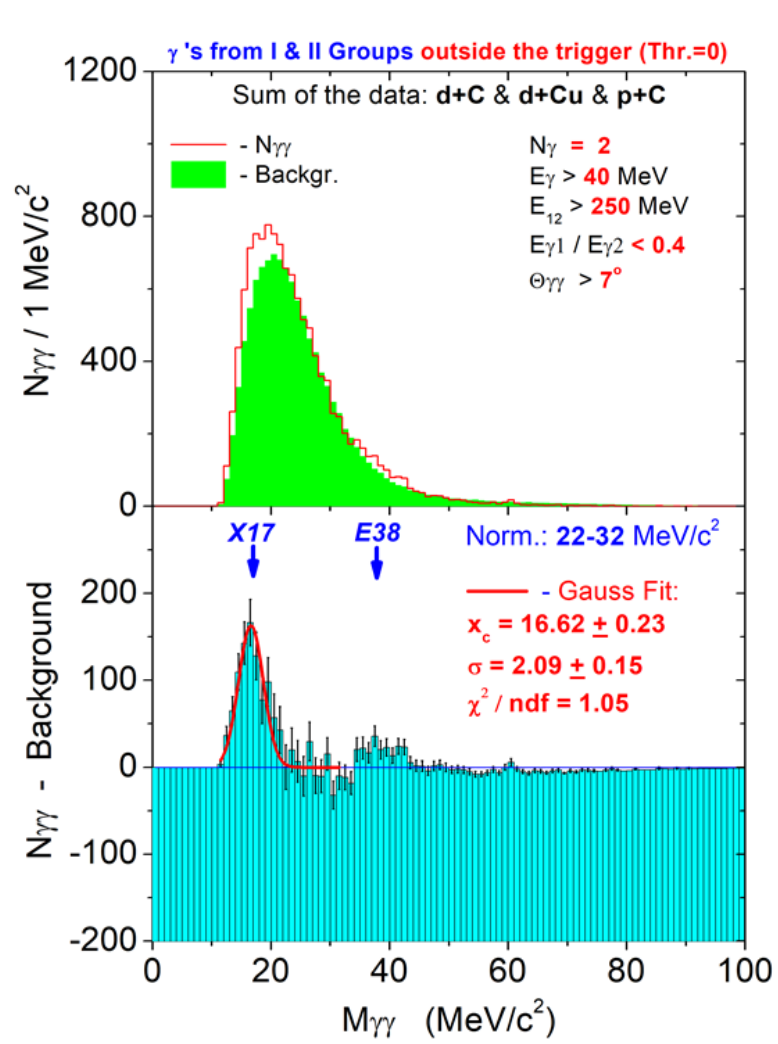}
\caption{ The diphoton invariant mass spectra from $d$ and $p$
  collisions with C and Cu targets at a few GeV per nucleon at the
  JINR Nuclotron, Dubna as given in \cite{Abr23}.  The solid curve in
  the upper panel shows the invariant mass distribution obtained by
  combining two photons from the same event, and green shaded region
  the invariant mass distribution by combining two photons from mixed
  events.  The signal of correlated photons subtracting the mixed
  event background gives the signal in the lower panel where the
  resonance-like structures at $\sim$17 and $\sim$38 MeV show up.  }
  \label{fig3}
\end{figure}

The observation of X17 and E38 at Dubna completes an important piece
of the anomalous particle puzzle as the isoscalar X17 and the
isovector E38 come in a pair, and they are orthogonal linear
combinations of the $|u\bar u\rangle $ and $|d \bar d\rangle$
components.   The signals for the E38 particle is however quite weak.
More measurements  need to be carried out to ensure that the E38 signals are genuine  and not 
experimental artifacts.

It is worth noting that if the two-photon decay of the X17 particle
observed at Dubna is confirmed, then the two-photon decay mode will
impose a constraint on the spin of the X17 particle on account of the
Landau-Yang theorem \cite{Lan48,Yan50}, which states that a massive
boson with spin 1 cannot decay into two photons.

The agreement of the X17 and E38  masses with those from 
phenomenological open-string model of $q\bar q$ QED mesons
\cite{Won10,Won20} lends support to the description that a quark and
an antiquark may be confined and bound as stable QED mesons
interacting in the Abelian U(1) QED interaction.  The confirmation of the
X17 and E38 particles will be therefore of great interest.

\section{The  $qqq$  color-singlet and  color-multiplet  quark matter}
\label{qqq}}

We have up to now considered only $q\bar q$ quark matter.  Our
consideration can be extended to $qqq$ quark matter in color space.
Quarks belong to the color-triplet $\bb{3}$ group.  From group
theoretical considerations, we have
\begin{eqnarray}
{\bb 3}\otimes {\bb 3}\otimes {\bb 3} = {\bb 1}
\oplus {\bb 8} \oplus {\bb 8} \oplus {\bb {10}}
\end{eqnarray}
 That is, the direct triple product of the triplet $\bb{3}$ group
 consists of the color-singlet $\bb{1}$ subgroup, two color-octet
 $\bb{8}$ subgroups , and one color-decuplet $\bb{10}$ subgroup.
 Therefore, three quarks combine to form the color-singlet $[q q q]^1$
 quark matter, two color-octet $[q qq]^8$ quark matters, and the
 color-decuplet $[q qq]^{10}$ quark matter.

By the principle of the colorless nature of observable entities,
observable quark matter complexes must be colorless color-singlet
entities.  Consequently, the color-octet and the color-decuplet $qqq$
quark matter must interact non-perturbatively first with the
color-octet SU(3) QCD interactions to form QCD-confined colorless
$[[qqq]^{8}g^8]^1$ or $[[qqq]^{ 10}g^8]^1$ complexes.  Additional QED
interaction in the color-multiplet $qqq$ quark matter can then be
included either perturbatively or nonperturbatively, and such an
additive inclusion does not change the colorless nature nor the
confinement property of these colorless complexes.  We therefore
recognize the $q q q$ color-octet and color-decuplet quark matter as
corresponding to the $q q q$ quark matter as conventionally envisaged
in the realm of our present knowledge.

On the other hand, the color-singlet $qqq$ quark matter is unexplored
and is now submitted for exploration.  Quarks in the color-singlet
$qqq$ quark matter must interact only with the color-singlet U(1) QED
interactions to form colorless color-singlet $[qqq]^1\gamma]^1$
  complexes for the lowest-energy region at $T$=0.  Inclusion of the
  SU(3) color-octet QCD interaction in the color-singlet $[qqq]^1$
  quark matter will change the colorless nature of these color-singlet
  complexes, or would place them at much higher energy excited states
  such as the $[[q q q]^1[gg]^1]^1$ states and the glueball states.
  
Of particular interest in the $qqq$ color-singlet quark matter is the
QED neutron with the $d$, $u$, and $d$
quarks~\cite{Won22,Won22a,Won22c}.  In such a color-singlet
$d$-$u$-$d$ system with three quarks of different colors, the
attractive QED interaction between the $u$ quark and the two $d$
quarks may overwhelm the repulsion between the two $d$ quarks and may
stabilize the QED neutron.  Upon examining the QED neutron in a
phenomenological three-body problem in 1+1 dimensions with an
effective interaction extracted from Schwinger’s exact QED solution in
1+1 dimensions, a phenomenological model in a variational calculation
yields a stable QED neutron at 44.5 MeV \cite{Won22,Won22a,Won22c}.
The analogous $u$$-$$d$$-$$u$ QED proton has been found to be
theoretically unstable because of the stronger QED repulsion between
the two $u$ quarks, and it does not provide a bound state nor a
continuum state for the QED neutron to decay onto, by way of the weak
interaction.  Hence, the QED neutron may be a dark neutron and may be
stable against the weak interaction.  It may have a very long lifetime
and may be a good candidate for the dark matter.  Because QED mesons
and QED neutrons may arise from the coalescence of deconfined quarks
during the deconfinement-to-confinement phase transition in different
environments, such as in high-energy heavy-ion collisions, neutron
star mergers~\cite{Bau19}, and neutron star cores~\cite{Ann20}, the
search for the QED bound states in various environments will be of
great interest.

It is interesting to explore the possibility of a QED neutron star as
a large assembly of QED neutrons.  The interaction between a QED
neutron and another QED neutron is through an attractive
electromagnetic polarization potential at large separations and a
repulsive Pauli-exclusion-type interaction at small separations.  The
electromagnetic interactions are quite weak.  A collection of a large
number of QED neutrons under their own gravitational field can be
stabilized by the attractive gravitational pressure and the repulsive
pressure of the degenerate QED neutron gas.  If a QCD neutron star and
a QED neutron star can be approximated as a polytropic gas spheres of
index $n=3$, as in the early consideration of Landau for a gravitating
and degenerate fermion gas \cite{Yak12,Lan32}, then from the
mathematical results of Landau, the maximum mass of a QCD or QED
neutron star is inversely proportional to the square of its
constituent mass.  One can thus estimate the maximum mass of the QED
neutron star and the maximum mass of the QCD neutron star to be given
approximately by
\begin{eqnarray}
\frac{M_{\rm max}(\text {QED neutron star})}
         {M_{\rm max}(\text {QCD neutron star})}
\sim \frac{(m_{\rm neutron}^\qcdu)^2}{(m_{\rm neutron}^\qedu)^2} .
\end{eqnarray}
Thus, for  $M_{\rm max}(\text {QCD neutron star}) \sim 1.6-2 $ $M_\odot$  \cite{Zel71} and 
$m_\qedd \sim 44.5 $ MeV \cite{Won22},  we have
\begin{eqnarray}
M_{\rm max}(\text {QED neutron star})   \sim 712- 890\, M_\odot.
\end{eqnarray}
Beyond such a maximum mass, an assembly of the QED neutrons will
collapse into a black hole.  It will be of great interest to explore
whether such QED neutron star or black hole may exist in the Universe.

\section{Implications on the possible existence of the QED mesons}

While the confirmation of the observations is pending, it is of
interest to examine the implication of the existence of the QED
mesons.  If the observations of the QED mesons are indeed confirmed
under further scrutiny, we will expect an expansion of our present
knowledge into new territories, including but not limited to the
following:

\begin{enumerate}

\item
 Color-singlet quark matter may exist, with possible neutral QED
 mesons at $T$=0.  There may then be the QED meson gas phase at $T<
 T_c ($QED), quark photon plasma above $T_c ($QED), and quark electron
 photon plasma at still higher temperatures.

\item
There may be a new family of QED-confined $q\bar q$ particles at $T$=0
that are composite in nature, with additional degrees of freedom in
spin-spin, spin-orbit, collective vibrations, collective rotations,
molecular states, …

\item
 Confinement occurs for $q$ and $\bar q$ not only in QCD but also in
 neutral color-singlet quark matter in QED. The group of
 quark-antiquark gauge interaction may be a broken U(3) group with
 U(3) = U(1) $\oplus$ SU(3).

\item
Confinement may be an intrinsic property of the quarks such that
quarks and antiquarks may interact with other different interactions,
including weak and gravitational interactions, to lead to confined
$q\bar q$ composite particles.

\item
The QED interaction between a quark and an antiquark may be predominantly
linear. In such a case, there may be a stable $d-u-d$ QED neutron, whereas
the corresponding $u-d-u$ QED proton is unstable. The QED neutron may be
a good candidate for dark matter, as we discussed in Section 5.

\item
 The confining QED interaction between a quark and an antiquark may
 differ from the non-confining QED interaction between an electron and
 a positron because of the additional color degrees of freedom of
 quarks.  The QED U(1) interaction between a quark and an antiquark is
 part of the greater but broken U(3) group of interactions in which
 U(3) = U(1) $\oplus$ SU(3), whereas the QED interaction between
 electron and positron is purely a U(1) interaction.  There is the
 question of whether the QED interaction between an electron and a
 positron may belong to the non-compact non-confining QED theory while
 the QED interaction between a quark and an antiquark belong to the
 confining compact QED theory.

\item
Astrophysical objects consisting of a large assembly of the isoscalar
QED mesons will be electron--positron emitters, gamma ray emitters, or
dark black holes with no emission depending on the mass of the
assembly.  Such assemblies of QED mesons present themselves as good
candidates of $e^+$$e^-$ emitters, gamma-ray emitters, or the
primordial dark matter \cite{Won20}.

\end{enumerate}

\section{Summary and discussions}

If the experimental QED mesons are confirmed, it would suggest that
the proposal of quarks interacting only in QED under appropriate
conditions would be a reasonable concept.  While we first presented
such an unusual and unfamiliar possibility as a phenomenology,
stimulated by Schwinger QED confinement mechanism and the observation
of the anomalous soft photons \cite{Won10}, we may now realize that it
may have a firmer and stronger foundation in the mathematical theory
of groups and the principle of colorless observable entities.  In the
process, we may expand our knowledge of the quark matter into the
uncharted frontiers in color space.

Quarks carry color and electric charges.  The direct product of the
quark color-triplet group and the antiquark color-antitriplet group
consists of the color-singlet subgroup and the color-octet subgroup.
Therefore, quarks and antiquarks combine to form the color-singlet
$q\bar q$ quark matter and the color-octet $q\bar q$ quark matter.
While the color-octet $q\bar q$ quark matter corresponds to the $q\bar
q$ quark matter commonly envisaged, the color-singlet $q\bar q$ quark
matter has been unexplored until now.

In order for the color-singlet quark matter to form color-singlet,
colorless complexes at $T\sim$ 0, the color-singlet quark matter must
interact only with the Abelian U(1) QED interaction while the
color-octet quark matter must interact in the non-Abelian SU(3) QCD
interaction.

Applying the Schwinger confinement mechanism to quarks with two light
flavors interacting only in QED, we find stable and confined $q\bar q$
isoscalar state at around 12 MeV and an isovector state around 33 MeV.
Including corrections due to the small quark masses with the
associated quark condensates leads to the isocalar QED meson mass at
about 17 MeV and the isovector meson mass at about 38 MeV.

The observations of the anomalous soft photons, the X17 particle, and
the E38 particle provide promising evidence for the possible existence
of the QED mesons confined and bound non-perturbatively by the QED
interaction.  Their masses at $\sim$17 and $\sim$38 MeV are close to
the theoretically predicted masses of isoscalar and isovector QED
mesons.  The occurrence of the isoscalar and isovector doublet
reflects properly the two-flavor nature of the light quarks.  Their
decays into $\gamma \gamma$ and $e^+e^-$ indicate their composite
nature and their connection to the QED interaction.  Their modes of
production by low-energy proton fusion and by high-energy nuclear
collisions can also be understood in terms of the production of
quark--antiquark pairs by soft gluon fusion or the $(q\bar q)$
production by string fragmentation in high-energy hadron--hadron
collisions~\cite{Won20,Won22,Won24}.

There remain many important questions which need to be experimentally
tested and resolved.  It has been conjectured all along that the
precursors of the anomalous soft photons may be the QED mesons with
masses of about 17 and 38 MeV \cite{Won10,Won20}.  A direct
experimental confirmation of such a conjecture for anomalous soft
photons remains lacking.  It will be of great interest in the new
study on the anomalous soft photons \cite{Bai24} to search for the QED
mesons.  The X17 and E38 particle have been uncovered at Dubna, and a
confirmation of these particles will be necessary to indicate the
two-flavor nature of the light quarks.  Experiments of the ATOMKI type
using the $p ^3$+H and $n$+$ ^3$He reactions may lead to stronger
$X17$ signals and may lessen the strong deformation effects present in
the internal pair conversion in the $^8$Be and $^{12}$C compound
nuclei.  New experimental measurements in search of the QED mesons
will bring us forward in the new physics frontier.

\vspace*{0.7cm}
\centerline{\bf Acknowledgments}
\vspace*{0.5cm} The author would like to thank Profs.\ T. Awes,
S. Sorensen, T. Cs\"org\H o, I. Y. Lee, Jack Yee Ng, C. M. Ko, and
S. A.  Chin for helpful discussions and communications.  The research
was supported in part by the Division of Nuclear Physics,
U.S. Department of Energy under Contract DE-AC05-00OR22725.


\vspace*{0.7cm}
\centerline{\bf References}

\begin{thebibliography}{9}                                                                                                %




\bibitem{Won93}
C.Y. Wong, {\it Introduction to High-Energy Heavy-Ion Collisions},
World Scientific Publishing, 1993.

\bibitem{Yag05} 
K. Yagi, T. Hatsuda, and Y. Miake, {\it Quark-Gluon Plasma}, Cambridge University Press, 2005.

\bibitem{Vog07}
R. Vogt, {\it Ultrarelativistic Heavy-Ion Collisions}, Elsevier Science, 2007.


\bibitem{Won10} 
 C. Y. Wong, {\it Anomalous soft photons in hadron production},
    Phys. Rev. C81, 064903 (2010), [arXiv:1001.1691].
    


\bibitem{Won20} 
C. Y. Wong, {\it Open string QED meson description of the X17
    particle and dark matter}, JHEP (2020) 165, [arxiv:2001.04864].
  

\bibitem{Per09} V. Perepelitsa, for the DELPHI Collaboration, {\it
  Anomalous soft photons in hadronic decays of Z$^0$}, Proceedings of
  the XXXIX International Symposium on Multiparticle Dynamics, Gomel,
  Belarus, September 4-9, 2009, Nonlin. Phenom. Complex Syst. 12, 343
  (2009).

\bibitem{Chl84} P.V. Chliapnikov $et~al.$, {\it Observation of direct
  soft photon production in $\pi^- p$ interactions at 280 GeV/c},
  Phys. Lett. { B 141}, 276 (1984).

\bibitem{Bot91} F. Botterweck $et~al.$ (EHS-NA22 Collaboration), {\it
  Direct soft photon production in $K^+p$ and $\pi^+p$ interactions at
  250 GeV/c}, Z. Phys. C 51, 541 (1991).

\bibitem{Ban93} S. Banerjee $et~al.$ (SOPHIE/WA83 Collaboration), {\it
  Observation of direct soft photon production in $\pi^- p$
  interactions at 280 GeV/c}, Phys. Lett. B 305, 182 (1993).

\bibitem{Bel97} A. Belogianni $et~ al.$ (WA91 Collaboration), {\it
  Confirmation of a soft photon signal in excess of QED expectations
  in $\pi^- p$ interactions at 280 GeV/c}, Phys. Lett. B 408, 487
  (1997).

\bibitem{Bel02pi}
 A. Belogianni $et~al.$ (WA102 Collaboration), {\it  Further analysis of a direct soft photon excess in pi- p interactions at 280-GeV/c},
 Phys. Lett.  {B548}, 122 (2002).


\bibitem{Bel02} A. Belogianni $et~ al.$ (WA102 Collaboration), {\it
  Observation of a soft photon signal in excess of QED expectations in
  $pp$ interactions}, Phys. Lett. B548, 129 (2002).
  

\bibitem{DEL06} J. Abdallah $et~al.$ (DELPHI Collaboration), {\it
  Evidence for an excess of soft photons in hadronic decays of Z$^0$}
  Eur.  Phys. J.  C47, 273 (2006), arXiv:hep-ex/0604038.

\bibitem{DEL08} J. Abdallah $et~al.$ (DELPHI Collaboration), {\it
  Observation of the muon inner bremsstrahlung at LEP1},
  Eur. Phys. J. C57, 499 (2008), arXiv:0901.4488.



\bibitem{DEL10} J. Abdallah $et~al.$ (DELPHI Collaboration), {\it
  Study of the dependence of direct soft photon production on the jet
  characteristics in hadronic Z$^0$ decays}, Eur. Phys. J. { C67}, 343
  (2010), arXiv:1004.1587.



\bibitem{Kra16}
 A. J. Krasznahorkay $et~al.$, {\it Observation of anomalous
    internal pair creation in $^8$Be: a possible indication of a
    light, neutral boson}, Phys. Rev. Lett. 116, 042501 (2016),
    [arXiv:1504.01527].


\bibitem{Kra19} A. J. Krasznahorkay $et~al.$, {\it New evidence
  supporting the existence of the hypothetical X17 particle},
  arXiv:1910.10459 (2019).



\bibitem{Nag19}
A. Nagy, A. J. Krasznahorkay, M. Ciemala, L. Csige, Z. Gacsi, M. Hunyadi, T. Klaus, M. Kmieck, A. Maj, N. Pietralla, Z. Revay, N. Sas, C. Stegorst, J. Timar, T. Tornyi, W. Wasilewska,  
{\it Searching for the double $\gamma$-decay of the X17 particle},
Nuo. Cim. 42C, 124  (2019).
   

\bibitem{Kra21} 
A. J. Krasznahorkay $et~al.$,
{\it New anomaly observed in $^4$He supports the existence of the hypothetical X17 particle},   Phys. Rev. C 104,   044003  (2021), [arxiv:2104.10075].



\bibitem{X17Kra}
A.J. Krasznahorkay,
{\it X17: status of the experiments on $^8$Be and  $^4$He},
Talk presented at the Workshop on ``Shedding Light on X17'', September 6, 2021,  Rome, Italy, in Ref.\ \cite{X1722}.

\bibitem{X1722}
Proceedings of the Workshop on ``Shedding Light on X17'',  September 6-8, 2021, Centro Ricerche Enrico Fermi, Rome, Italy; 
Editors: M. Raggi,  P. Valente, M. Nardecchia,  A. Frankenthal,  G. Cavoto,
published in D. S. M. Alves $et~al.$,  Eur. Phys. J. C 83,  230 (2023). 


\bibitem{Sas22}
N.J. Sas, A.J. Krasznahorkay, M. Csatl\' os, J. Guly\' as, B. Kert\' esz, A. Krasznahorkay, J. Moln\' ar, I. Rajta, J. Tim\' ar, I. Vajda, M.N. Harakeh,
{\it 
Observation of the X17 anomaly in the 7Li(p,$e^+e^-$)8Be direct proton-capture reaction
},
[arXiv:2205.07744].


\bibitem{Kra22}
A. J. Krasznahorkay $et~al.$,
{\it 
New anomaly observed in $^{12}$C supports the existence and the vector character of the hypothetical X17 boson
},  [arXiv:2209.10795].


\bibitem{Kra23}
A.J. Krasznahorkay $et~al.$,
{\it Observation of the X17 anomaly in the decay of the Giant Dipole Resonance of $^8$Be},
Talk presented at the 
International Symposium on Multiparticle Dynamics,
at Gy\" ongy\" os, Hungary, August 20-26, 2023,  arXiv:2308.06473.





\bibitem{Abr23}
K.  Abraamyan, Ch. Austin, M.I. Baznat, K.K. Gudima, M.A. Kozhin, S.G. Reznikov, A.S. Sorin,
{\it Observation of structures at $\sim$17 and $\sim$38 MeV/c2 in the $\gamma \gamma$  invariant mass spectra in pC, dC, and dCu collisions at plab of a few GeV/c per nucleon},
   Physics of particle and Nuclei, 55(4), 868 (2024), arxiv:.2311.18632.



\bibitem{Abr12}
K. Abraamyan, A. B. Anisimov, M. I. Baznat, K .K. Gudima,
M. A. Nazarenko, S. G. Reznikov, and A.S. Sorin,
{\it Observation of the E(38)-boson},
arxiv:1208.3829v1 (2012).

\bibitem{Abr19}
K. Abraamyan, C. Austin, M. Baznat, K. Gudima,
M. Kozhin, S. Reznikov, and A. Sorin,
{\it Check of the structure in photon pairs spectra
at the invariant mass of about 38 {\rm MeV/}$c^2$},
E PJ Web of Conferences 204, 08004 (2019).


\bibitem{Abr09}
K. U. Abraamyan $et ~al.$, 
{\it Resonance structure in the $\gamma\gamma$ 
invariant mass spectrum in
$p$C and
$d$C interactions},
Phys. Rev. C80, 034001 (2009).


\bibitem{Tra24}
The-Anh Tran $et~al.$,
{\it Confirmation the $^8${\rm Be} anomaly
with a different spectrometer},   Universe  10(4), 168 (2024), arxiv:2401.11676

\bibitem{Sch62}
J. Schwinger,
{\it Gauge invariance and mass  II},
Phys. Rev. 128, 2425 (1962).

\bibitem{Sch63}
J. Schwinger,
{\it Gauge theory of vector particles},
in Theoretical Physics, Trieste Lectures, 1962 (IAEA, Vienna,
1963), p. 89.




\bibitem{Won09}
C. Y.  Wong,
{\it The Wigner function of produced particles in string fragmentation},
Phys. Rev. C80, 054917 (2009),  arXiv:0903.3879.






  \bibitem{Won22} 
C. Y. Wong, {\it On the stability of the open-string QED neutron
    and dark matter}, Euro. Phys. Jour. A 58, 100 (2022), [arxiv:2010.13948].  


 \bibitem{Won22a} 
C. Y. Wong, {\it 
    QED mesons, the QED neutron, and the dark matter}, in 
 Proceedings of the    19th  International Conference on Strangeness in Quark Matter,  EPJ Web of Conferences 259, 13016 (2022), [arXiv:2108.00959].
    

 \bibitem{Won22c} 
C. Y. Wong, {\it On the question of  quark confinement in the  QED  interaction}, Front. of Phys, 18, 64401 (2023),  [arxiv:2208.09920].


 \bibitem{Won23} 
 C. Y. Wong and A. Koshelkin,
{\it Dynamics of quarks and gauge fields in the lowest-energy states in QCD and QED},
Euro. Phys. J. A 59, 285 (2023), 
[arXiv:2111.14933].


\bibitem{Won24}
C. Y. Wong,  Talk Presented at 52th International Symposium on Multiparticle Dynamics, August 21-25, 2023, Gy\" ongy\" os, Hungary, published in Universe (2024), 10, 173, arxiv:2401.04142.

  


\bibitem{And83}
B. Andersson, G. Gustafson, and T. Sj\"ostrand,
{\it A general model for jet fragmentation},
  Zeit. f{\"u}r Phys. {   C20}, 317 (1983).


\bibitem{Cos17}
L. Cosmai, P. Cea, F. Cuteri, A. Papa,
{\it Flux tubes in QCD with (2+1) HISQ fermions},
Pos, 4th annual International Symposium on Lattice Field Theory
24-30 July 2016
University of Southampton, UK,
arxiv:1701.03371 (2017).


\bibitem{Gel68}
M. Gell-Mann, R.J. Oakes and B. Renner, Behavior of current divergences under
SU(3) × SU(3), Phys. Rev. 175 (1968) 2195.


\bibitem{Pes95}
M. E. Peskin, D. V. Schroeder, {\it An Introduction to Quantum
Field Theory}, Addison-Wesley Publishing Company, 1995.


\bibitem{Ros50}
M. E. Rose, {\it Internal Pair Formation},
Phys. Rev. 78, 184 (1950).

\bibitem{Mc76}
K. McDonald, 
{\it Physics Examples
and other Pedagogic Diversions, Neutral-Pion Decay}, Joseph Henry Laboratories, Princeton University, Princeton, NJ 08544
(September 15, 1976; updated June 4, 2019), 
URL: http://kirkmcd.princeton.edu/examples/piondecay.pdf.



\bibitem{Bar23}
D. Barducci and C. Toni,
{\it An updated view on the ATOMKI nuclear anomalies},
arxiv:2212.06453.



\bibitem{Fen16} J. Feng $et~al.$, {\it Protophobic fifth force
  interpretation of the observed anomaly in $^8${\rm Be} nuclear transitions},
  Phys. Rev. Lett. 2016 117, 071803 (2016); 
  
  \bibitem{Fen17}
  J. Feng $et~ al.$, {\it
    Particle physics models for the 17 MeV anomaly in beryllium
    nuclear decays}, Phys. Rev. D 95, 035017 (2017).

\bibitem{Fen20}
J. L. Feng, J. M. P. Tait, and B. Verhaaren,
{\it Dynamical Evidence For a Fifth Force Explanation
of the ATOMKI Nuclear Anomalies},
Phys. Rev. D 102, 036016 (2020),
arxiv:2006.01151.

\bibitem{Var24}
S. Varr\'o,  Talk Presented at 52th International Symposium on Multiparticle Dynamics, August 21-25, 2023, Gy\" ongy\" os, Hungary, published in 
Universe 10, 86 (2024).

\bibitem{Zha17} Xilin Zhang, G. A Miller, {\it Can nuclear physics
  explain the anomaly observed in the internal pair production in the
  Beryllium-8 nucleus? }, Phys. Lett. B773, 159 (2017),
  arXiv:1703.04588.

\bibitem{Hay22}
A. C. Hayes, J. Friar , G. M. Hale , and G. T. Garvey,
{\it
Angular correlations in the  $e^+ e^-$  decay of excited states in $^8$Be
},
Phys. Rev. C 105, 055502 (2022).

 
\bibitem{PAD25}
F. Bossi $et~al.$,
{\it Search for a new 17 MeV resonance via $e^+ e^-$  annihilation with the PADME Experiment},
arXiv:2505.24797.

\bibitem{MEG25}
K. Afanaciev et al. (MEG II),
{\it Search for the X17 particle in $^7$Li(p,$e^+e^-)^8$Be processes with the MEG II detector},
 (2024), arXiv:2411.07994
.


\bibitem{Bev11}
E. van Beveren and G. Rupp, {\it First indications of the existence of a 38 MeV light scalar boson}
arxiv:1102.1863 (2011).


\bibitem{Lan48}
L. D. Landau, {\it The moment of a 2-photon system},
Dold. Akcad. Nauk. Ser. Fiz. 60, 207 (1948).

\bibitem{Yan50}
C. N. Yang, {\it 
 Selection Rules for the Dematerialization of a Particle into Two Photons},
Phys. Rev. 77, 242 (1950).




\bibitem{Bau19}
A. Bauswein, N-U. F. Bastian, D.  Blaschke, K. Chatziioannou, 
J. A. Clark, T. Fischer, and M. Oertel,
{\it Identifying a first-order phase transition in neutron-star
mergers through gravitational waves},
Phys. Rev. Lett.122, 061102 (2019).

\bibitem{Ann20}
E. Annala, T. Gorda, A. Kurkela,
J. Naettilae, and A. Vuorinen, 
{\it Evidence for quark-matter cores in massive
neutron stars},
Nat. Phys.  16, 907 (2020)





\bibitem{Bai24}
R. Bailhache $et~al.$,   
{\it Anomalous soft photons: status and perspectives},  Phys.  Rept. 1097, 1 (2024),
arXiv:2406.17959.


\bibitem{Yak12}
D. Yakovlev, P. Haensel, G. Baym, and C. Tethick, arix:1210.0682.

\bibitem{Lan32}
L. D. Landau, Phys. Z. Sowjetunion 1, 285 (1932).

\bibitem{Zel71}
Ya. B. Zeldovich and I. D. Novikov, {\it Relativistic Astrophysics Vol. I}, Unversity of Chicago Press, 1971, p. 283.




\end{thebibliography}
\end{document}